\documentclass[conference, letterpaper]{IEEEtran}

\IEEEoverridecommandlockouts

\usepackage{cite}
\usepackage{url}
\usepackage{fancyvrb}
\usepackage[final]{pdfpages}
\usepackage{longtable}
\usepackage{booktabs}
\usepackage{amsmath}
\usepackage{amssymb}
\usepackage{makecell}
\usepackage{soul}
\usepackage{graphicx}
\usepackage{algorithm}
\usepackage{algorithmic}

\usepackage{txfonts}
\DeclareRobustCommand{\IEEEauthorrefmark}[1]{\smash{\textsuperscript{\footnotesize #1}}}

\usepackage{url}
\usepackage[hidelinks]{hyperref}

\usepackage{etoolbox}

\begin{document}

\include{header}
\title{\huge Null-Space Flow Matching for MIMO Channel \\ Estimation in Latency-Constrained Systems}

\author{
\IEEEauthorblockN{Junjie Zhao\IEEEauthorrefmark{1}, Guangming Liang\IEEEauthorrefmark{2,*}, Xiaonan Liu\IEEEauthorrefmark{3}, Dongzhu Liu\IEEEauthorrefmark{2}}
\IEEEauthorblockA{\IEEEauthorrefmark{1}College of Electronic Information and Optical Engineering, Nankai University}
\IEEEauthorblockA{\IEEEauthorrefmark{2}School of Computing Science, University of Glasgow}
\IEEEauthorblockA{\IEEEauthorrefmark{3}School of Natural and Computing Science, University of Aberdeen}
\IEEEauthorblockA{\IEEEauthorrefmark{*}Guangming Liang is the Corresponding Author}
\IEEEauthorblockA{Emails: 2313495@mail.nankai.edu.cn, g.liang.1@research.gla.ac.uk, xiaonan.liu@abdn.ac.uk, dongzhu.liu@glasgow.ac.uk}
}

\maketitle

\begin{abstract}
Accurate yet low-latency channel state information (CSI) acquisition is essential for multiple-input multiple-output (MIMO) communication systems. While advanced deep generative models, such as score-based and diffusion models, enable high-fidelity CSI reconstruction from limited pilot observations, they often suffer from high inference latency. To achieve accurate CSI estimation under stringent latency constraints, this paper proposes a null-space flow matching (FM) framework that leverages a range-null space decomposition to separate observation-informed and underdetermined channel components. Specifically, the pilot observations are used to regulate the observable range-space channel component, while an FM-based generative prior primarily resolves the ambiguous null-space degrees of freedom through iterative refinement. To further improve the robustness and efficiency of the proposed framework, we introduce a noise-aware adaptive correction strategy to suppress channel noise on the refinement trajectory, along with a power-law time schedule to better allocate the limited number of refinement steps. Experimental results demonstrate that our method achieves competitive normalized mean square error (NMSE) performance even under a strict latency budget of around 3~ms, while delivering a superior accuracy-latency tradeoff compared with both model-based and generative baselines.
\end{abstract}


\section{Introduction}

Accurate channel state information (CSI) acquisition is a fundamental enabler for sixth-generation (6G) networks, as it underpins key functionalities such as beamforming, adaptive transmission, and radio resource management \cite{10054381}. However, acquiring high-resolution CSI is fundamentally constrained in the spatial and temporal domains. On one hand, the adoption of massive multiple-input multiple-output (MIMO) architectures and the increasing density of user access substantially enlarge the spatial dimension of wireless channels, thereby increasing the pilot overhead \cite{10845870}. On the other hand, the proliferation of high-mobility scenarios, such as high-speed railways and unmanned aerial vehicles, introduces significant Doppler shifts that drastically shorten the channel coherence time, leaving only a narrow temporal window for pilot transmission, CSI estimation, and subsequent link adaptation \cite{9832933}. Therefore, acquiring highly accurate CSI with limited pilot overhead and low latency remains a major bottleneck in 6G communications.

Generative models have attracted increasing attention for reconstructing high-fidelity CSI from limited pilot observations, as they can capture channel distributions and provide informative priors. Classical Gaussian mixture models (GMMs) \cite{9842343} approximate channel distributions as a weighted sum of Gaussian components and exploit their mean vectors and covariance matrices for linear minimum mean square error (LMMSE) estimation \cite{kay1993fundamentals}, but the fixed parametric form often mismatches real channels in complex propagation environments. Deep generative models relax this distributional assumption by employing neural networks to learn channel distributions. Among them, early variational autoencoders (VAEs) \cite{10629241} map the pilot observations into a latent representation, from which the first- and second-order moments are inferred for LMMSE channel estimators. However, the latent representation is regularized toward a simple prior distribution, hindering the preservation of fine-grained channel characteristics. 

To address this issue, advanced score-based models \cite{arvinteMIMOChannelEstimation2023a} and diffusion models \cite{zhouGenerativeDiffusionModels2025a} operate directly in the original channel space to better capture physical scattering structures and spatial correlations. Specifically, score-based models estimate the score function and iteratively guide noisy samples toward high-probability regions of the channel distribution, whereas diffusion models progressively recover clean channel samples by reversing a predefined noise-injection process through iterative denoising. Nevertheless, the performance gain comes at the expense of extensive iterative refinement, whose latency may exceed the millisecond-scale channel coherence time\footnote{For instance, at a carrier frequency of 2~GHz and a user speed of 90~km/h, the channel coherence time is around 3~ms.}, making the estimated CSI outdated for downstream tasks \cite{10417075}. Consequently, the key challenge is to learn high-quality generative priors while producing CSI estimates that remain faithful to pilot observations within the coherence-time budget.

% To address this issue, score-based models \cite{arvinteMIMOChannelEstimation2023a} learn the score function and iteratively steer noisy samples toward high-probability channel regions, while diffusion models \cite{zhouGenerativeDiffusionModels2025a} learn to progressively recover clean channel samples by reversing a predefined noise-injection process through iterative denoising. By operating directly in the original channel space, both approaches can better capture physical scattering structures and spatial correlations than latent-variable models. Nevertheless, the performance gain comes at the expense of extensive iterative refinement, whose latency may exceed the millisecond-scale channel coherence time\footnote{For instance, at a carrier frequency of 2~GHz and a user speed of 90~km/h, the channel coherence time is around 3~ms.}, making the estimated CSI outdated for downstream tasks \cite{10417075}. Consequently, the key challenge is to learn high-quality generative priors while producing CSI estimates that remain faithful to pilot observations within the stringent coherence-time budget.

% and diffusion models \cite{zhouGenerativeDiffusionModels2025a} to learn priors directly in the original channel space, which is better suited to modeling the physical scattering structures and spatial correlations. 

Recently, flow matching (FM) \cite{lipmanFLOWMATCHINGGENERATIVE2023a} has emerged as a promising generative paradigm that achieves a favorable balance between generation quality and inference speed. In contrast to score-based and diffusion models that refine samples via stochastic dynamics and often follow highly curved trajectories, FM learns a deterministic velocity field that transports samples along straighter paths \cite{liu2023flow}. This geometric reformulation allows larger inference step sizes without substantially increasing discretization error, thereby reducing the iteration number while maintaining high generation fidelity. Such an efficiency advantage makes FM particularly appealing for real-time channel estimation, and several studies \cite{liu2025flowmatchingbasedgenerativemodels,jiangRecursiveFlowGenerative2026,zhangFLOWChannelEstimationMIMO2026} have exploited FM as a generative prior over wireless channels and coupled it with observation-informed iterative refinement to recover CSI using limited pilots. 

Despite this progress, existing FM-based estimators typically perform generative refinement in the full channel space without exploiting the measurement-induced range-null channel structure. Since the pilot observations directly constrain the range-space channel component, treating all channel components uniformly during iterative inference may lead to unnecessary updates of already observation-informed directions, rather than explicitly concentrating the refinement on the unresolved channel ambiguity. This may reduce inference efficiency and make it more difficult to remain faithful to the pilot observations under a limited refinement budget. 

To overcome the limitations, we propose a null-space flow matching framework with the main contributions as follows:

% However, existing FM-based estimators still generate the full-space channel from scratch, even though pilot measurements can be used to directly reconstruct the range-space channel component and leave only the null-space component underconstrained \cite{wangZEROSHOTIMAGERESTORATION,wickremasingheFlowSteerConditioningFlow2025}. This implies that they solve a higher-dimensional generation problem than necessary, since a nontrivial portion of the iterative refinement is wasted on re-estimating the already-determined content instead of focusing on the truly uncertain null-space degrees of freedom. As a result, the misallocation of generative modeling capacity not only compromises the inference efficiency of the estimator, but also weakens its ability to remain faithful to the pilot observations.

\noindent $\bullet$ \textbf{Null-Space Flow Matching for Channel Estimation:} 
We first employ FM to learn the channel prior by modeling a continuous transformation from a standard Gaussian distribution to the empirical channel distribution. We then formulate CSI estimation as an iterative null-space refinement process, in which the observable range-space channel component is constrained by the pilot observation, while the ambiguous null-space channel component is progressively refined under the guidance of the learned FM-based generative prior.

\noindent $\bullet$ \textbf{Noise-Aware Correction and Non-Uniform Time Schedule:} To enhance the robustness and efficiency of our proposed framework, we introduce a noise-aware adaptive correction mechanism to mitigate the tendency of observation noise to drive the final estimate away from the learned channel manifold, and develop a power-law time schedule to allocate denser steps to the early refinement stage, where the estimate is more vulnerable to discretization errors.

\noindent $\bullet$ \textbf{Superior Estimation Accuracy and Inference Speed:} 
Experimental results demonstrate that our proposed method achieves a superior accuracy-latency tradeoff, with a normalized mean square error (NMSE) below $-15$~dB within a tight coherence-time budget of around $3$~ms.  Moreover, it consistently yields the lowest NMSE among the compared model-based and generative schemes over the evaluated latency range, lying on the empirical Pareto frontier of estimation accuracy and inference latency.

\section{System Model and Preliminaries}
% This section introduces the system model for MIMO channel estimation in Sec.~II-A, and presents the FM-based generative prior in Sec.~II-B as preliminaries to the proposed method.

\subsection{MIMO Channel Estimation}
We consider a point-to-point narrowband MIMO system, where the transmitter and receiver are equipped with $N_t$ and $N_r$ antennas, respectively. The CSI between the transmitter and receiver is denoted by $\mathbf{H} \in \mathbb{C}^{N_r \times N_t}$, which is assumed to remain constant during the channel estimation period. A total of $N_p$ pilot symbols are selected from the first $N_p$ columns of a $N_t \times N_t$ normalized Hadamard matrix, and the $k$-th pilot symbol is denoted by $\mathbf{p}_k \in \mathbb{C}^{N_t \times 1}$. During the pilot transmission phase, the received signal at the $k$-th symbol is given by
\begin{equation}\label{eq:received_signal}
   \mathbf{y}_k = \mathbf{H}\mathbf{p}_k + \mathbf{n}_k,
\end{equation}
where $\mathbf{n}_k \sim \mathcal{CN}\!\left(\mathbf{0}, 2\sigma_n^2 \mathbf{I}\right)$ denotes the additive white Gaussian noise vector, with variance $\sigma_n^2$ for both the real and imaginary parts. By stacking (\ref{eq:received_signal}) over $N_p$ pilot symbols, the received signal is written in the matrix form as
\begin{equation}
    \mathbf{Y} = \mathbf{H}\mathbf{P} + \mathbf{N},
\end{equation}
where $\mathbf{Y} = [\mathbf{y}_1, \ldots, \mathbf{y}_{N_p}]$, $\mathbf{P} = [\mathbf{p}_1, \ldots, \mathbf{p}_{N_p}]$, and $\mathbf{N} = [\mathbf{n}_1, \ldots, \mathbf{n}_{N_p}]$. Using vectorization, we reformulate the signal model as
\begin{equation}\label{eq:measurement_equation_0}
    \bar{\mathbf{y}} = \left(\mathbf{P}^T \otimes \mathbf{I}_{N_r}\right)  \bar{\mathbf{h}} + \bar{\mathbf{n}},
\end{equation}
where $\bar{\mathbf{y}} = \mathrm{vec}(\mathbf{Y}) \in \mathbb{C}^{N_r N_p \times 1}$, $\bar{\mathbf{h}} = \mathrm{vec}(\mathbf{H}) \in \mathbb{C}^{N_r N_t \times 1}$, $\bar{\mathbf{n}} = \mathrm{vec}(\mathbf{N}) \in \mathbb{C}^{N_r N_p \times 1}$, and $\otimes$ represents Kronecker product. 

Due to the limited number of propagation clusters and the narrow angular spread, wireless channels at high-frequency bands often exhibit sparsity in the angular domain. Motivated by this phenomenon and following \cite{zhouGenerativeDiffusionModels2025a,liang2025environmentawarechannelinferencecrossmodal}, we perform channel estimation on the angular-domain representation $\bar{\mathbf{h}}_{\mathrm{ad}}$, which is related to $\bar{\mathbf{h}}$ by
\begin{equation}\label{eq:angular_representation}
\bar{\mathbf{h}} = \left( (\mathbf{A}_T^{T})^{H} \otimes \mathbf{A}_R \right)\bar{\mathbf{h}}_{\mathrm{ad}},
\end{equation}
where $\mathbf{A}_T \in \mathbb{C}^{N_t \times N_t}$ and $\mathbf{A}_R \in \mathbb{C}^{N_r \times N_r}$ denote the discrete Fourier transform matrices corresponding to the array response matrices at the transmitter and receiver, respectively. We then substitute \eqref{eq:angular_representation} into \eqref{eq:measurement_equation_0} and obtain
\begin{equation}\label{eq:measurement_equation_1}
\bar{\mathbf{y}} = \bar{\mathbf{A}}_{\mathrm{ad}} \bar{\mathbf{h}}_{\mathrm{ad}} + \bar{\mathbf{n}},
\end{equation}
where $\bar{\mathbf{A}}_{\mathrm{ad}} = \left(\mathbf{P}^T \otimes \mathbf{I}_{N_r}\right) \left( (\mathbf{A}_T^{T})^{H} \otimes \mathbf{A}_R \right) \in \mathbb{C}^{N_r N_p \times N_r N_t}$. Since neural networks operate on real-valued tensors, we convert \eqref{eq:measurement_equation_1} into an equivalent real-valued form by concatenating the real and imaginary parts:
\begin{equation}\label{eq:measurement_equation_2}
\mathbf{y} = \mathbf{A}\mathbf{h} + \mathbf{n},
\end{equation}
where
\begin{equation}
\mathbf{y}=
\begin{bmatrix}
\Re\{\bar{\mathbf{y}}\}\\
\Im\{\bar{\mathbf{y}}\}
\end{bmatrix}
\in \mathbb{R}^{M \times 1},
\mathbf{h}=
\begin{bmatrix}
\Re\{\bar{\mathbf{h}}_{\mathrm{ad}}\}\\
\Im\{\bar{\mathbf{h}}_{\mathrm{ad}}\}
\end{bmatrix}
\in \mathbb{R}^{N \times 1},
\mathbf{n}=
\begin{bmatrix}
\Re\{\bar{\mathbf{n}}\}\\
\Im\{\bar{\mathbf{n}}\}
\end{bmatrix}
\in \mathbb{R}^{M \times 1},
\end{equation}
and
\begin{equation}
\mathbf{A}=
\begin{bmatrix}
\Re\{\bar{\mathbf{A}}_{\mathrm{ad}}\} & -\Im\{\bar{\mathbf{A}}_{\mathrm{ad}}\}\\
\Im\{\bar{\mathbf{A}}_{\mathrm{ad}}\} & \Re\{\bar{\mathbf{A}}_{\mathrm{ad}}\}
\end{bmatrix}
\in \mathbb{R}^{M \times N},
\end{equation}
with $M = 2N_r N_p$ and $N = 2N_r N_t$, and $\Re(\cdot)$ and $\Im(\cdot)$ denoting the real and imaginary parts of a complex vector, respectively.

% \begin{figure*} [t]
%     \centering
%     \includesvg[width=1\linewidth] {frame} 
%     \caption{Schematic of the proposed null-space flow matching method for MIMO channel estimation}
%     \label{fig:placeholder}
% \end{figure*} 

\subsection{FM-based Generative Prior}\label{sec:FM_prior}
A generative prior is a channel prior distribution defined explicitly or implicitly by generative models, which constrain the candidate channel estimates to lie near the manifold of realistic channel realizations. Unlike handcrafted priors that rely on simplified structural assumptions (e.g., sparsity and low rank), the generative prior is learned in a data-driven manner and thus can capture more complex channel statistics. 

In this work, we employ FM to model the generative channel prior as a continuous-time transport process that evolves an initial state drawn from a standard Gaussian distribution (i.e., $\mathbf{h}_0 \sim p_0$) to a terminal state following the empirical channel distribution (i.e., $\mathbf{h}_1 \sim p_1$). To characterize the evolution, we introduce the intermediate state $\mathbf{h}_t \sim p_t$ at time $t\in[0,1]$, where $\{p_t\}_{t\in[0,1]}$ represents a probability path connecting $p_0$ and $p_1$. This path is governed by a time-dependent velocity field $u(\mathbf{h}_t,t)$ that satisfies the ordinary differential equation (ODE)
\begin{align}
	\frac{d\mathbf{h}_t}{dt} = u(\mathbf{h}_t,t).
	\label{eq:ODE_definition_a}
\end{align}
Following \cite{lipmanFLOWMATCHINGGENERATIVE2023a,liang2025environmentawarechannelinferencecrossmodal}, we define the evolution process using a linear interpolation path, i.e.,
\begin{align}\label{eq:intermediate_state}
	\mathbf{h}_t = (1-t)\mathbf{h}_0 + t\mathbf{h}_1,
\end{align}
while the target velocity is obtained by differentiating (\ref{eq:intermediate_state}) with respect to $t$, yielding
\begin{align}
	\hat{u}(\mathbf{h}_t,t)= \mathbf{h}_1-\mathbf{h}_0.
\end{align}
Based on this formulation, we employ a time-dependent neural network $v_\theta(\mathbf{h}_t,t)$ parameterized by $\theta$ to approximate the target velocity field. The network is trained by minimizing the mean squared error between the predicted and target velocities:
\begin{align}
	\min_{\theta}~\mathbb{E}_{\substack{t \sim \mathcal{U}[0,1], \mathbf{h}_0\sim p_0,\mathbf{h}_1\sim p_1}}
	\left\|v_\theta\big((1-t)\mathbf{h}_0+t\mathbf{h}_1,t\big)-(\mathbf{h}_1-\mathbf{h}_0)\right\|_F^2.
	\label{eq:FlowMatching}
\end{align}
After training, channels are generated by first drawing $\mathbf{h}_0 \sim p_0$ and then integrating the learned ODE from $t=0$ to $t=1$.

\section{Null-Space Flow Matching for MIMO \\ Channel Estimation}
% In this section, we propose a null-space flow matching method to achieve high-accuracy and low-latency MIMO channel estimation, including the problem formulation in Sec.~III-A and the FM-based channel estimation in Sec.~III-B.

\subsection{Problem Formulation}
The objective of MIMO channel estimation is to recover $\mathbf{h}$ from the received pilot observation $\mathbf{y}$ given the measurement matrix $\mathbf{A}$. Let $\alpha = N_p / N_t$ denote the pilot density. When $\alpha < 1$, the channel estimation becomes an underdetermined linear inverse problem, in which the measurement equation (\ref{eq:measurement_equation_2}) admits infinitely many solutions. To regularize this ill-posed problem, we seek an estimate $\hat{\mathbf{h}}\in \mathbb{R}^{N \times 1}$ that balances the following two desirable properties:
\begin{equation}
\mathit{Observation~Fidelity:}~ \mathbf{A}\hat{\mathbf{h}} \approx  \mathbf{y},
\quad
\mathit{Channel~Realism:}~ \hat{\mathbf{h}} \sim p_1,
\end{equation}
where the observation-fidelity criterion encourages compatibility with the received pilots, while the realism criterion favors solutions aligned with the learned channel prior. Since the FM model in Sec.~\ref{sec:FM_prior} captures the channel distribution, it can naturally serve as a generative prior for the latter criterion.

To explicitly characterize the ambiguity induced by the underdetermined measurement model, we adopt the range-null space decomposition. Specifically, by introducing the pseudo-inverse of the measurement matrix as $\mathbf{A}^{\dagger} =\mathbf{A}^T(\mathbf{AA}^T)^{-1}$, we decompose the realistic channel $\mathbf{h}$ into a range-space component and a null-space component with respect to $\mathbf{A}$:
\begin{equation}\label{eq:Space_Decomposition}
    \mathbf{h} = \underbrace{\mathbf{A}^\dagger \mathbf{A} \mathbf{h}}_{\mathbf{h}_{\text{range}}} + \underbrace{(\mathbf{I} - \mathbf{A}^\dagger \mathbf{A})\mathbf{h}}_{\mathbf{h}_{\text{null}}}.
\end{equation}
Using the pseudo-inverse property $\mathbf{A}\mathbf{A}^{\dagger}\mathbf{A} \equiv \mathbf{A}$, the range-space component satisfies $\mathbf{A}\mathbf{h}_{\text{range}} \equiv \mathbf{A}\mathbf{A}^{\dagger}\mathbf{A}\mathbf{h} \equiv \mathbf{A}\mathbf{h}$, while the null-space component satisfies $\mathbf{A}\mathbf{h}_{\text{null}} \equiv \mathbf{A}(\mathbf{I} - \mathbf{A}^{\dagger}\mathbf{A})\mathbf{h} \equiv 0$. These relations show that the pilot measurement $\mathbf{y}=\mathbf{A}\mathbf{h}+\mathbf{n}$ constrains only the range-space channel component, while the null-space component remains invisible to $\mathbf{A}$.

Inspired by denoising diffusion null-space model (DDNM) \cite{wangZEROSHOTIMAGERESTORATION} and FlowSteer \cite{wickremasingheFlowSteerConditioningFlow2025}, which were originally developed for image restoration, we adapt their underlying ideas to channel estimation and construct the channel estimate $\hat{\mathbf{h}}$ based on the range-null space decomposition in (\ref{eq:Space_Decomposition}):
\begin{equation}\label{eq:estimation_strategy}
    \hat{\mathbf{h}} = \mathbf{A}^{\dagger}\mathbf{y} + (\mathbf{I} - \mathbf{A}^{\dagger}\mathbf{A})\tilde{\mathbf{h}},
\end{equation}
where $\mathbf{A}^{\dagger}\mathbf{y}$ is the range-space channel component recovered from pilot measurements, and $(\mathbf{I} - \mathbf{A}^{\dagger}\mathbf{A})\tilde{\mathbf{h}}$ represents the null-space channel component induced by the FM-based generative prior. Consequently, our goal is to exploit the pre-trained FM model to generate a suitable $\tilde{\mathbf{h}}$, whose null-space projection is in harmony with the observable range-space component, thereby yielding a final channel estimate $\hat{\mathbf{h}}$ that balances observation fidelity and channel realism.

\subsection{Iterative Null-Space Refinement for Channel Estimation}
We cast the CSI estimation as an iterative null-space refinement process that evolves an initial state $\mathbf{h}_0 \sim \mathcal{N}(0, \mathbf{I})$ toward a terminal estimate $\mathbf{h}_1$ that balances observation fidelity and channel realism. To this end, we perform $K$ iterations of a predict-then-correct procedure, where each iteration first applies an FM prediction to move the current state toward regions favored by the learned channel prior, followed by a pilot-guided correction to promote observation fidelity.

Formally, we define time grids as $0=t_0<t_1<\cdots<t_K=1$. Given the current state $\mathbf{h}_{t_k}$, we first apply an Euler step along the learned velocity field $v_{\theta}(\cdot,t)$:
\begin{equation}\label{eq:iter_predict}
    \tilde{\mathbf{h}}_{t_{k+1}}=\mathbf{h}_{t_k}+v_{\theta}(\mathbf{h}_{t_k},t_k)(t_{k+1}-t_k).
\end{equation}
This FM prediction step improves compatibility with the learned channel prior but does not explicitly incorporate the pilot measurement. Then, according to (\ref{eq:estimation_strategy}), we consider a hard-correction step that keeps the range-space part fixed to the observation-anchored term $\mathbf{A}^{\dagger}\mathbf{y}$, and retains only the null-space component of the FM prediction. The resulting correction step is expressed as
\begin{align}
    \mathbf{h}_{t_{k+1}}=\mathbf{A}^{\dagger}\mathbf{y}+(\mathbf{I}-\mathbf{A}^{\dagger}\mathbf{A})\tilde{\mathbf{h}}_{t_{k+1}},
\end{align}
which can equivalently be written as
\begin{align}\label{eq:iter_project}
    \mathbf{h}_{t_{k+1}}=\tilde{\mathbf h}_{t_{k+1}}-\mathbf A^{\dagger}\bigl(\mathbf A\tilde{\mathbf h}_{t_{k+1}}-\mathbf y\bigr),
\end{align}
where $\mathbf A^{\dagger}\bigl(\mathbf A\tilde{\mathbf h}_{t_{k+1}}-\mathbf y\bigr)$ is viewed as an observation correction term applied to the FM prediction $\tilde{\mathbf h}_{t_{k+1}}$. This hard-correction form exactly reproduces the received pilot observation:
\begin{equation}\label{eq:iter_consistency}
    \mathbf{A}\mathbf{h}_{t_{k+1}}=\mathbf{A}\mathbf{A}^{\dagger}\mathbf{y}+\mathbf{A}(\mathbf{I}-\mathbf{A}^{\dagger}\mathbf{A})\tilde{\mathbf{h}}_{t_{k+1}}=\mathbf{A}\mathbf{A}^{\dagger}\mathbf{y}+\mathbf{0}=\mathbf{y}.
\end{equation}
Thus, the hard-correction step (\ref{eq:iter_project}) corresponds to exact observation fidelity. 

Despite ensuring exact observation fidelity, the above hard-correction scheme does not explicitly account for measurement noise and may therefore degrade channel realism. Moreover, the commonly used uniform time discretization may be suboptimal under a limited refinement budget. We therefore introduce two additional designs to enhance the robustness and efficiency of the iterative inference process.

\begin{algorithm}[t]
\caption{Null-Space FM-based Channel Estimation}
\begin{algorithmic}[1]\label{alg:Infernce_scheme}
\REQUIRE Pre-trained velocity field $v_{\theta}(\cdot,t)$, measurement matrix $\mathbf{A}$, pilot observation $\mathbf{y}$, number of inference steps $K$, power-law index $\rho$, and observation noise level $\sigma_n$.
\STATE Compute pseudo-inverse $\mathbf{A}^\dagger = \mathbf{A}^T(\mathbf{AA}^T)^{-1}$.
\STATE Initialize the power-law time schedule $\{t_k\}_{k=0}^{K}$ via (\ref{eq:Non-Uniform_Schedule}) and sample the initial state as $\mathbf{h}_0 \sim \mathcal{N}(0, \mathbf{I})$.
\FOR{$k = 0$ \TO $K-1$}
    \STATE Obtain the FM prediction $\tilde{\mathbf{h}}_{t_{k+1}}$ via (\ref{eq:iter_predict}).
    \STATE Calculate the guidance factor $\eta(t_{k+1})$ via (\ref{eq:gudiance_factor}).
    \STATE Obtain the corrected state $\mathbf{h}_{t_{k+1}}$ via (\ref{eq:soft_correction}).
\ENDFOR
\RETURN Estimated channel vector $\hat{\mathbf{h}} = \mathbf{h}_1$.
\end{algorithmic}
\end{algorithm}

\subsubsection{Noise-Aware Adaptive Correction}
In (\ref{eq:iter_project}), the pilot observation $\mathbf{y}=\mathbf{A}\mathbf{h}+\mathbf{n}$ is contaminated by channel noise. Therefore, enforcing hard correction throughout the inference process may inject measurement noise into the refinement trajectory and push the final estimate away from the learned channel manifold. 

% Intuitively, we should impose a strong observation correction when the estimate is uncertain, and weaken it as the estimate becomes more reliable.

% To balance observation fidelity and robustness to measurement noise, we impose a strong observation correction when the estimate is uncertain, and weaken it as the estimate becomes more reliable.
% In (\ref{eq:iter_project}), the range-space approximation $\mathbf{A}^{\dagger}\mathbf{y}=\mathbf{A}^{\dagger}\mathbf{A}\mathbf{h}+\mathbf{A}^{\dagger}\mathbf{n}$ is contaminated by channel noise. Therefore, enforcing constant observation correction throughout the whole inference may inject observation noise into the generative trajectory and push the estimate away from the channel manifold. Intuitively, we should impose a strong observation correction when the estimate is uncertain, and weaken it as the estimate becomes more reliable.

To balance observation fidelity and channel realism in the presence of measurement noise, we introduce a time-dependent guidance factor based on the relative reliability of the FM prediction and the pilot observation. For the linear interpolation path (\ref{eq:intermediate_state}) used in FM, the randomness inherited from Gaussian initialization scales with $1-t$ and vanishes as $t\to 1$. Hence, $1-t_{k+1}$ serves as a proxy for the uncertainty of $\tilde{\mathbf h}_{t_{k+1}}$, while $\sigma_n$ represents the noise level. Their ratio measures the benefit of observation correction relative to the risk of noise injection, yielding
\begin{equation}
\eta(t_{k+1})
\propto
\frac{1-t_{k+1}}{\sigma_n}.
\end{equation}
To prevent over-correction, we clip the factor at unity:
\begin{equation}\label{eq:gudiance_factor}
\eta(t_{k+1})
=
\min\!\left(1.0,\frac{1-t_{k+1}}{\sigma_n}\right).
\end{equation}
Accordingly, the hard correction in (\ref{eq:iter_project}) is replaced by the following adaptive correction:
\begin{equation}\label{eq:soft_correction}
    \mathbf h_{t_{k+1}}=\tilde{\mathbf h}_{t_{k+1}}-\eta(t_{k+1})\,\mathbf A^{\dagger}\bigl(\mathbf A\tilde{\mathbf h}_{t_{k+1}}-\mathbf y\bigr).
\end{equation}

\subsubsection{Power-Law Time Schedule}
Standard FM inference typically adopts a uniform time discretization scheme, i.e., $t_k=k/K$, where $k=0,\ldots,K$. However, under a limited number of refinement steps imposed by latency constraints, uniform discretization can degrade estimation quality because the difficulty of refinement varies over time. In the early stage (i.e., $t\to 0$), the estimate is still dominated by Gaussian initialization and remains far from the target channel manifold, which makes the update more sensitive to discretization error and thus requires finer steps. As inference proceeds, the estimate moves closer to the channel manifold, and updates become more consistent across adjacent times, thereby allowing larger step sizes to improve efficiency.

To better allocate the limited refinement budget, we replace the uniform time grid with a power-law schedule:
\begin{equation}\label{eq:Non-Uniform_Schedule}
    t_k = \left(\frac{k}{K}\right)^{\rho},~\forall k\in\{0,1,\ldots,K\},
\end{equation}
where $\rho>1$ assigns denser steps to the early stage and progressively larger steps to the later stage.

We refer to this approach as \emph{Null-Space Flow Matching}, since the range-null space decomposition enables the FM-based generative prior to focus primarily on the underdetermined null-space structure, while the observable range-space component is regulated by the pilot measurement during inference. The complete generative refinement process to yield the final channel estimate $\hat{\mathbf{h}}=\mathbf{h}_1$ is summarized in \textbf{Algorithm~\ref{alg:Infernce_scheme}}.

\begin{table}[t]
    \centering
    \caption{Inference Latency of Different Schemes}
    \label{tab:latency}
    \begin{tabular}{|l|c|c|}
    \hline
    \textbf{Method} & \textbf{Preprocess Latency [ms]} & \textbf{Latency per Step [ms]} \\ 
    \hline
    Vanilla-LMMSE  &  1.763 & 5.761   \\ 
    \hline
    GMM-LMMSE    & 66.624   & 241.320   \\ 
    \hline
    VAE-LMMSE     & 9.077 & 48.827    \\ 
    \hline
    SGM     & 64.662 & 4.615    \\ 
    \hline
    DPS     & 1.83 & 0.142    \\ 
    \hline
    FM-PGD  & 4.13 & 0.139     \\ 
    \hline
    Our method & 2.58 & 0.132 \\ 
    \hline
    \end{tabular}
\end{table}

\section{Experiments}
In this section, we evaluate the performance of the proposed null-space flow matching framework using 3GPP TR 38.901-compliant channels \cite{3gpp.38.901}. Specifically, we simulate a MIMO system where the transmitter and receiver are equipped with $N_t = 16$ and $N_r = 64$ antennas, respectively. The MATLAB 5G Toolbox \cite{matlab_5g_toolbox} is used to generate a balanced mixture of CDL-A, CDL-B, CDL-C, and CDL-D channels. The dataset contains a total of $20{,}000$ channel samples $\mathbf{H} \in \mathbb{C}^{64 \times 16}$, with all four profiles uniformly mixed in both the training and testing partitions. The samples are normalized to satisfy $\mathbb{E}[\lvert{\mathbf{H}}_{ij}\rvert^2] = 1$, and then divided into $80\%$ for training and $20\%$ for testing. The pilot observation $\mathbf{y}$ is generated according to the signal model in (\ref{eq:measurement_equation_2}) across a wide range of signal-to-noise ratio (SNR) defined as $N_t/(2\sigma_n^2)$. 

The proposed null-space flow matching approach is implemented in PyTorch. The neural velocity field $v_\theta(\cdot,t)$ is built upon the time-dependent U-Net architecture following \cite{liang2025environmentawarechannelinferencecrossmodal}, which is trained and tested on an NVIDIA GeForce RTX 5090 32GB GPU.  During the inference, we select a power-law index of $\rho=1.5$ using hyperparameter tuning. The available number of inference steps $K$ is determined by the coherence-time budget, which is calculated by 
\begin{equation}\label{eq:K_calculation}
    K= \max\Big(0,\lfloor\frac{\text{Time Budget~-~Preprocess Latency}}{\text{Latency per Step}}\rfloor\Big),
\end{equation}
where the \emph{Preprocess Latency} accounts for the time consumed by the initial operations before entering the main inference loop (e.g., the computing time of the pseudo-inverse $\mathbf{A}^{\dagger}$ in our method), and the \emph{Latency per Step} corresponds to the execution time of a single iteration.

% Per-sample inference and per-step latencies are measured on the same NVIDIA GeForce RTX 5090 with batch size one and averaged over 100 test samples. CUDA synchronization is applied immediately before and after each timed inference and refinement step to account for asynchronous GPU execution; for conservative evaluation, the preprocessing latency includes the computation of $\mathbf{A}^{\dagger}$, although it can be precomputed when the pilot matrix is fixed.

\begin{table}[t]
    \centering
    \caption{Ablation Studies at $\alpha = 0.625$ and $K = 5$ in NMSE ($\mathrm{dB}$)}
    \label{tab:ablation}
    \scriptsize
    \resizebox{\columnwidth}{!}{%
    \begin{tabular}{|c|c|c|c|c|c|c|c|}
    \hline
    \makecell{\textbf{Noise-Aware}\\\textbf{Correction}} & \makecell{\textbf{Power-Law}\\\textbf{Schedule}} &  \textbf{$-5$ dB} & \textbf{0 dB} & \textbf{5 dB} & \textbf{10 dB} & \textbf{15 dB} & \textbf{20 dB} \\
    \hline
    $\times$ & $\times$ & $-9.8$ & $-13.5$ & $-16.8$ & $-18.6$ & $-21.1$ & $-22.4$ \\
    \hline
    $\times$ & $ \checkmark$ & $-10.5$ & $-14.4$ & $-17.7$ & $-19.3$ & $-21.9$ & $-23.1$ \\
    \hline
    $\checkmark $ & $\times$ & $-12.4$ & $-16.4$ & $-19.0$ & $-20.0$ & $-22.5$ & $-23.4$ \\
    \hline
    $\checkmark$ & $\checkmark$ & $\mathbf{-13.2}$ & $\mathbf{-17.2}$ & $\mathbf{-19.9}$ & $\mathbf{-20.6}$ & $\mathbf{-23.1}$ & $\mathbf{-24.0}$ \\
    \hline
    \end{tabular}
        }
\end{table}

\subsection{Baselines and Metric}
We compare the proposed null-space flow matching method against six representative benchmarks. Note that the first three baselines are the single-step methods that produce the channel estimate in a single evaluation step, while the last three baselines and our proposed method are iterative approaches that require repeated function evaluations during inference. Their inference latencies are reported in Table~\ref{tab:latency}.

\noindent $\bullet$ \textbf{Vanilla-LMMSE \cite{kay1993fundamentals}}. 
The channel is estimated using a vanilla LMMSE estimator, where the first- and second-order statistics are estimated offline from historical samples.

\noindent $\bullet$ \textbf{GMM-LMMSE \cite{9842343}}. 
The channel distribution is modeled as a weighted sum of fifteen Gaussian components, each with its own mean vector and covariance matrix. An LMMSE estimate is computed for each component, and the final channel estimate is obtained by combining these estimates using their posterior probabilities.

\noindent $\bullet$ \textbf{VAE-LMMSE \cite{10629241}}. 
The coarse least-squares estimate from pilot observations is encoded into a latent representation, from which the conditional first- and second-order CSI moments are inferred and then used for LMMSE estimation.

\noindent $\bullet$ \textbf{Score-Based Generative Model (SGM) \cite{arvinteMIMOChannelEstimation2023a}}. 
The score function of the channel distribution is learned via score matching, and the channel estimation is performed via posterior inference with Langevin dynamics.

\noindent $\bullet$ \textbf{Diffusion Posterior Sampling (DPS) \cite{zhouGenerativeDiffusionModels2025a}}. 
The channel prior is learned using a denoising diffusion probabilistic model, and the channel is estimated via posterior sampling with pilot observation-guided gradients.

\noindent $\bullet$ \textbf{FM with Proximal Gradient Descent (FM-PGD) \cite{zhangFLOWChannelEstimationMIMO2026}}. 
The same FM-based generative prior as ours is used. However, unlike in our method, the channel estimation is performed in the full channel space, with the generative prior integrated into a proximal gradient descent framework.

The channel estimation accuracy is quantified by averaging the per-sample linear NMSE over the complete testing set $\mathcal{T}$ and then converting the result to decibels:
\begin{equation}
\mathrm{NMSE~[dB]} = 10 \log_{10}
\left(
\frac{1}{|\mathcal{T}|}\sum_{i\in\mathcal{T}}
\frac{\|\hat{\mathbf{h}}^{(i)} - \mathbf{h}^{(i)}\|_2^2}
{\|\mathbf{h}^{(i)}\|_2^2}
\right),
\end{equation}
where $\|\cdot\|_2$ represents the $\ell_2$-norm. %Thus, the per-sample NMSE values are first averaged in the linear domain, and the logarithm is applied only after this averaging, rather than averaging per-sample NMSE values in dB.

\subsection{Experimental Results}
\subsubsection{Ablation Studies}
In Table~\ref{tab:ablation}, we conduct ablation studies to evaluate the impact of the power-law schedule and noise-aware correction independently at pilot density $\alpha=0.625$ and $K=5$. Both components consistently improve the NMSE across different SNRs, while their combination achieves the best performance in all cases. In particular, the noise-aware correction yields gains of $2.6$ and $2.9$ dB at $-5$ and $0$ dB SNR, respectively, indicating its effectiveness under strong observation noise. In contrast, the power-law schedule provides a stable improvement of about $0.7$--$0.9$ dB across the entire SNR range under the 5-step refinement budget. Combining the two components further improves the NMSE by up to $3.7$ dB compared with the baseline without either component.

\begin{figure}[t]
    \centering
    \includegraphics[width=1\linewidth]{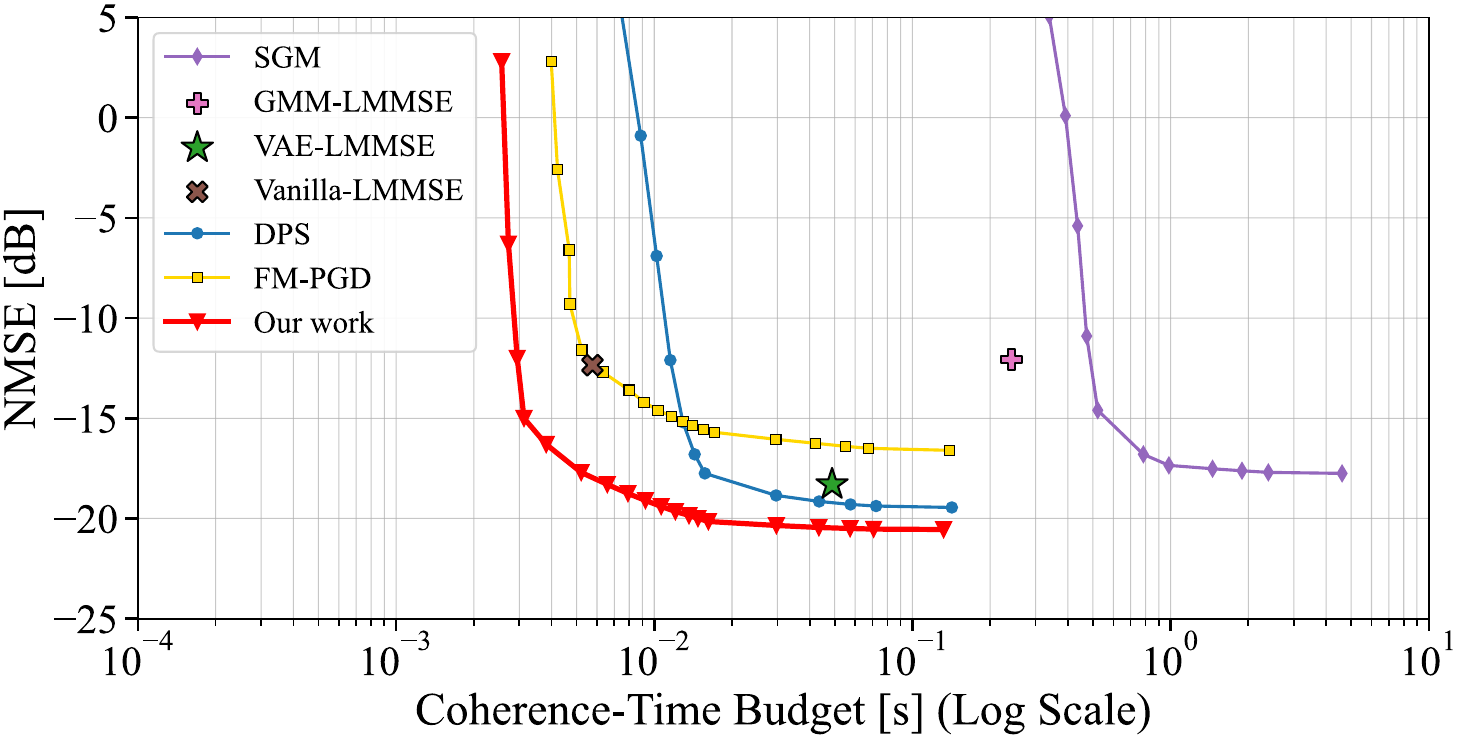}
    \caption{The impact of coherent-time budget on NMSE performance under SNR=10~dB and pilot density $\alpha$=0.625.}
    \label{fig:latency}
\end{figure}

\begin{figure}[t]
    \centering
    \includegraphics[width=1\linewidth]{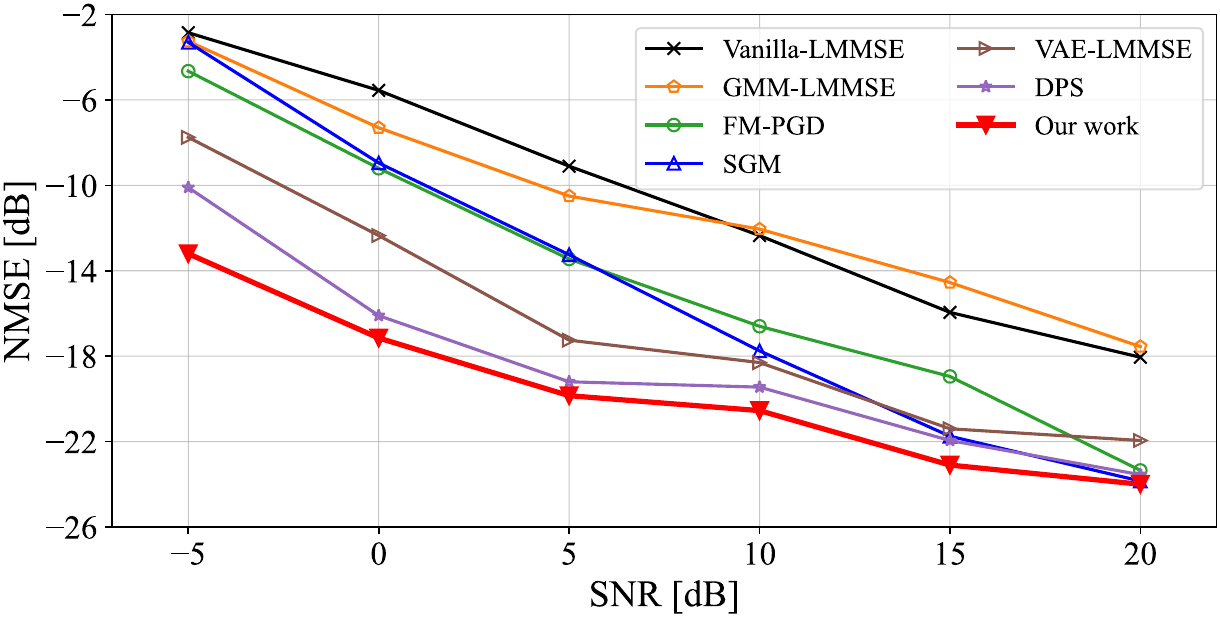}
    \caption{The impact of SNR on NMSE performance under pilot density $\alpha$=0.625 and sufficient coherence-time budget.}
    \label{fig:snr}
\end{figure}

\begin{figure}[t]
    \centering
    \includegraphics[width=1\linewidth]{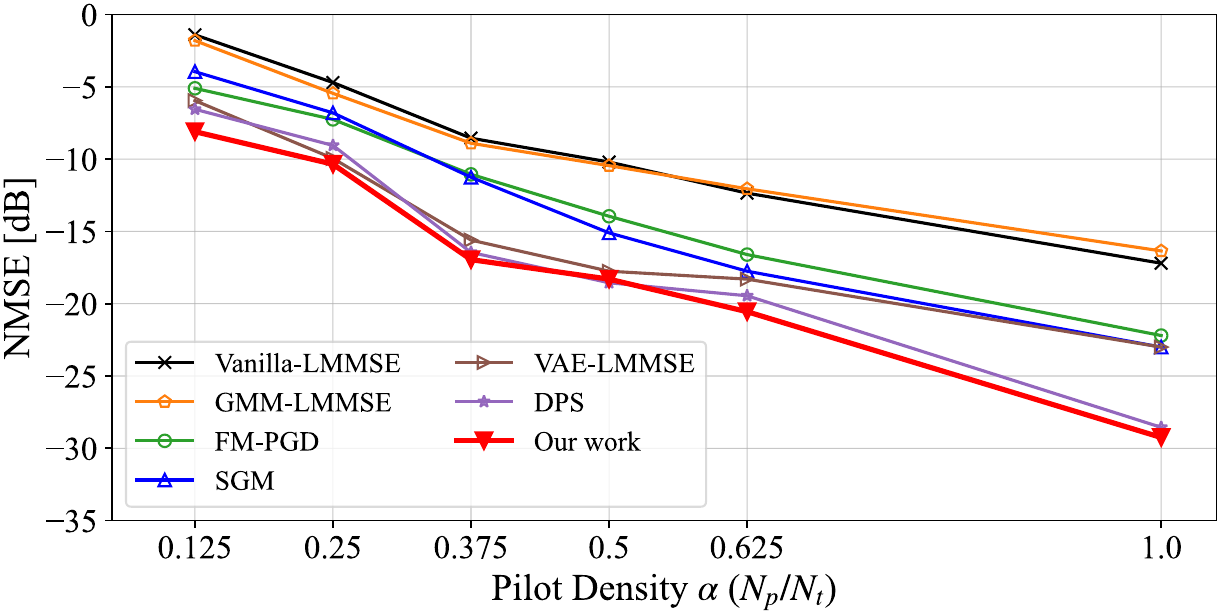}
    \caption{The impact of pilot density on NMSE performance under SNR=10~dB and sufficient coherence-time budget.}
    \label{fig:cr}
\end{figure}

\subsubsection{Accuracy versus Coherence-Time Budget}
Fig.~\ref{fig:latency} illustrates the NMSE performance across varying coherence-time budgets. In this experiment, the  SNR is fixed at 10~dB, the pilot density is set to $\alpha$ = 0.625 and the number of inference steps $K$ is calculated according to Table~\ref{tab:latency} and (\ref{eq:K_calculation}). It is observed that the proposed method achieves competitive NMSE even under extremely tight time budgets, reaching below $-15$~dB when the coherence-time budget is around $3\times10^{-3}$~s and further improving to about $-20$~dB as the time budget increases. More importantly, our method lies on the Pareto front of the accuracy-latency tradeoff, consistently achieving the lowest NMSE across all schemes over the evaluated latency range.

By contrast, the iterative baselines are more sensitive to the available time budget. Specifically, DPS suffers a sharp performance degradation when the budget falls below $10^{-2}$~s. The SGM requires a substantially larger budget, on the order of $10^{-1}$-$10^{0}$~s, to approach a competitive NMSE. This behavior is mainly attributed to the curved sampling trajectories of the score-based and diffusion models, which require many discretization steps to fully exploit their inference capability. Although the FM-PGD benefits from the straighter transport path of flow matching, it exhibits an inferior accuracy-latency tradeoff compared to our method. This is because it generates the full-space channel from scratch and does not explicitly exploit the pilot-induced range-space component. Furthermore, while the single-step baselines have relatively low latency, their estimation accuracy is fundamentally limited by the restricted expressive power of channel priors.

\subsubsection{Accuracy under Sufficient Coherence-Time Budget}
To examine the performance limit of the iterative generative methods under ample computation, we fix the number of inference steps to $K=1000$ in Fig.~\ref{fig:snr} and Fig.~\ref{fig:cr}. %This ensures that each iterative method is given a sufficient coherence-time budget to approach its performance limit.

Fig.~\ref{fig:snr} studies the impact of SNR on NMSE performance under pilot density $\alpha$ = 0.625. It is observed that the proposed method achieves the lowest NMSE among all compared methods over the entire SNR range. This result highlights the effectiveness of the proposed noise-aware adaptive correction mechanism, especially in low-SNR regimes where the pilot observations are heavily corrupted by channel noise.

Fig. \ref{fig:cr} investigates the impact of pilot density $\alpha$ on NMSE performance under SNR = 10~dB. It is observed that the proposed method achieves the lowest NMSE over most of the evaluated pilot-density range. Even at a very low pilot density of $\alpha = 0.125$, our method attains $-8.10$~dB. This advantage stems from the division of roles between the range and null spaces, which provides a favorable balance between observation fidelity and learned channel realism under highly underdetermined conditions.

\section{Conclusion}
This work proposes a null-space flow matching framework for accurate channel estimation in latency-constrained MIMO systems. By leveraging range-null space decomposition, the proposed method uses pilot-informed corrections to regulate the observable range-space channel component while exploiting the FM prior primarily for the underdetermined null-space structure. We further design a noise-aware adaptive correction mechanism to balance observation fidelity and channel realism in the presence of measurement noise, and introduce a power-law time schedule to allocate limited refinement steps more efficiently. Experimental results demonstrate favorable NMSE performance under stringent coherence-time constraints.

\bibliographystyle{IEEEtran} 
\bibliography{GLOBAL}

\end{document}